\documentclass{nature} 
\usepackage[utf8]{inputenc}
\usepackage[english]{babel}
\usepackage{enumerate}
\usepackage{color}

\usepackage{dcolumn}
\usepackage{bm}
\usepackage{mathptmx}
\usepackage{pifont}
\usepackage{cancel}
\usepackage{amssymb}
\usepackage{amsmath}
\usepackage{upgreek}
\usepackage{xcolor}
\usepackage{graphicx}
\usepackage[normalem]{ulem}
\newcommand\bluesout{\bgroup\markoverwith{\textcolor{blue}{\rule[0.5ex]{2pt}{1.1pt}}}\ULon}
\makeatletter
\let\saved@includegraphics\includegraphics
\AtBeginDocument{\let\includegraphics\saved@includegraphics}
\renewenvironment*{figure}{\@float{figure}}{\end@float}
\makeatother
\title{The interplay of crystal-field transitions and  exchange spin dynamics in a ferrimagnet}

\author{Arpita Dutta,$^{1,2}$ Pratyay Mukherjee,$^{3}$ Ritwik Mondal,$^{3}$ and Shovon Pal$^{1,2}$}

\begin{document}
\maketitle
\begin{affiliations}
\item School of Physical Sciences, National Institute of Science Education and Research, Jatni, 752 050 Odisha, India
\item Homi Bhaba National Institute, Training School Complex, Anushakti Nagar, 400 094 Mumbai, India
\item Department of Physics, Indian Institute of Technology (ISM) Dhanbad, 826 004 Dhanbad, India
\end{affiliations}

\maketitle

\begin{abstract}
Rare-earth iron garnets offer an ideal platform for exploring the interplay of low-energy excitations and the complex temperature-dependent magnetization dynamics. In these systems, exchange coupling between rare-earth and iron sublattices generates high-frequency collective spin excitations. In addition, the robust spin-orbit coupling of localized 4$f$ electrons triggers the crystal-electric-field (CEF) transitions at THz frequencies. Despite extensive research into the garnet spin dynamics, the interplay between CEF excitations and exchange modes has remained largely unmapped. Using temperature-dependent THz time-domain spectroscopy, we demonstrate a hybridization between the Yb-ion CEF excitation and the Yb-Fe exchange mode in Gd$_{3/2}$Yb$_{1/2}$BiFe$_{5}$O$_{12}$. This coupling is characterized by a significant redistribution of spectral and temporal weights as the material approaches its magnetization compensation temperature. Notably, the Yb-Fe exchange mode exhibits an anomalous redshift upon cooling -- a reversal of the conventional blue shift typically driven by increased exchange coupling. We trace this phenomenon to a modification of Yb-Fe exchange anisotropy, driven by the interplay of the Fe exchange field and Yb CEF excitations. These findings highlight the critical role of CEF-mediated exchange coupling in shaping low-energy spin dynamics, positioning rare-earth garnets as a cornerstone for future THz spintronic technologies.
\end{abstract}

\section*{INTRODUCTION}
Manipulating spin excitations at THz frequencies, where magnetic resonances oscillate on picosecond timescales, is a foundational challenge for the next generation of ultrafast spintronic technologies. THz-frequency spin dynamics enables switching speeds far beyond those of conventional ferromagnetic devices and underpins emerging applications in high-density magnetic memory and magnon-based signal processing. For the implementation of such technologies, materials should exhibit strong inter-sublattice exchange interactions that give rise to THz-frequency resonance modes and enable tuning of their frequencies over a broad temperature range, extending to relatively high temperatures. Ferrimagnets occupy a uniquely advantageous position in this landscape. Unlike ferromagnets, whose resonance frequencies are limited to the GHz frequency range by weak anisotropy fields, ferrimagnets support antiferromagnet-like exchange dynamics that push collective spin excitations into the THz regime~\cite{Kampfrath2011, BaierlPRL2016, Blank2021THz, Baltz2018, Jungwirth2016} while retaining a finite net magnetization~\cite{Kim2022} that simplifies experimental access to the excitation and detection of the exchange mode.

This unique architecture enables ferrimagnets to support both exchange-dominated THz dynamics and significant magneto-optical and spintronic effects~\cite{HoffmannPRA2015}. Furthermore, their temperature-dependent multi-sublattice interactions provide a tunable platform for exploring ultrafast spin phenomena, viz., nonequilibrium spin transfer, spin–orbit coupling effects, and laser-induced angular momentum exchange between sublattices~\cite{Radu2011, Stanciu2007, ostler12, Kirilyuk2010, Kimel2005}. In this context, rare-earth iron garnets are particularly compelling ferrimagnetic systems because of their strong interplay between iron (Fe) 3$d$ and rare-earth (RE) 4$f$ moments~\cite{Nakamoto2017, Neel1964, Calhuon1957}. The exchange interaction between them gives rise to collective spin excitation in the THz frequency range, whose dynamics are governed by strong exchange coupling, anisotropy fields, the applied external magnetic field, and temperature. A defining feature of these garnets is the distinct thermal evolution of their sublattice magnetizations, which gives rise to compensation points~\cite{Bai2025, MashkovichPRB2022}. At the magnetization compensation temperature ($T_{\rm MC}$), the opposing RE and Fe moments cancel out with strongly reduced stray fields, allowing the material to mimic the robust properties of an antiferromagnet~\cite{VivianePRB2022, HaltzPRB2022, Stanciu2006PRB, Gomonay2018, Donges2017, Srinivasan2024}. The near-$T_{\rm MC}$ regime is particularly transformative for spintronics; it enables efficient spin control and high-speed switching driven by time-dependent magnetic fields, effectively bypassing the speed limits of conventional ferromagnetic systems~\cite{DebPRB2018, Blank2021THz}. Note that an angular momentum compensation temperature also exists for ferrimagnets, where the total angular momentum of the sublattices compensates due to their different gyromagnetic ratios. Such a regime is not addressed here, as our study is primarily concerned with the magnetic excitation spectrum near vanishing net magnetization~\cite{Kim2017, Mentink2012}.

Rare-earth iron garnets exhibit a fascinating characteristic, namely, the crystal-electric-field (CEF) excitations of the rare-earth ions that reside in the THz frequency range~\cite{LeendersPRL2025, BaierlNP2016}. In these systems, the electrostatic potential from the surrounding ligand ions lifts the degeneracy of the rare-earth 4$f$ orbitals. This results in crystal-field-split energy levels with separations resonant to THz frequencies. At low temperatures, these CEF levels gain prominence; due to the strong spin-orbit coupling of the localized 4$f$ electrons, the magnetic anisotropy is modified by CEF excitation, which eventually reshapes the low-energy spin dynamics~\cite{BaierlNP2016}. Consequently, localized CEF modes hybridize with collective exchange interactions, leaving a distinct fingerprint through complex coupled dynamics. However, experimental investigations into this interplay -- particularly near the compensation point, where the net magnetization is suppressed -- remain largely unexplored.

Using temperature-dependent THz time-domain spectroscopy (THz-TDS), we investigate the interplay between Yb-ion CEF transitions and sublattice-dependent exchange dynamics in the ferrimagnetic Gd$_{3/2}$Yb$_{1/2}$BiFe$_{5}$O$_{12}$ (GdYb-BIG) garnet across its magnetic compensation temperature ($T_{\rm MC}$). We observe hybridization between these low-energy excitations, accompanied by a pronounced redistribution of temporal and spectral weight near the compensation point. The distinct orbital angular momenta of the rare-earth ions (Gd and Yb), coupled with their unique sublattice magnetization profiles, generate a temperature-dependent effective magnetic field driven by strong Fe exchange interactions. Notably, the Yb-Fe exchange mode exhibits an anomalous redshift upon cooling—a stark departure from the conventional blue-shift typically triggered by enhanced exchange coupling at low temperatures. We attribute this anomaly to a modification of the anisotropic Yb-Fe exchange energy, arising from the synergy of the Fe exchange field and Yb-ion CEF excitations, mediated by robust spin-orbit coupling.

\section*{RESULTS AND DISCUSSION}
\begin{figure}
 \centering
 \includegraphics[width=1.0\textwidth]{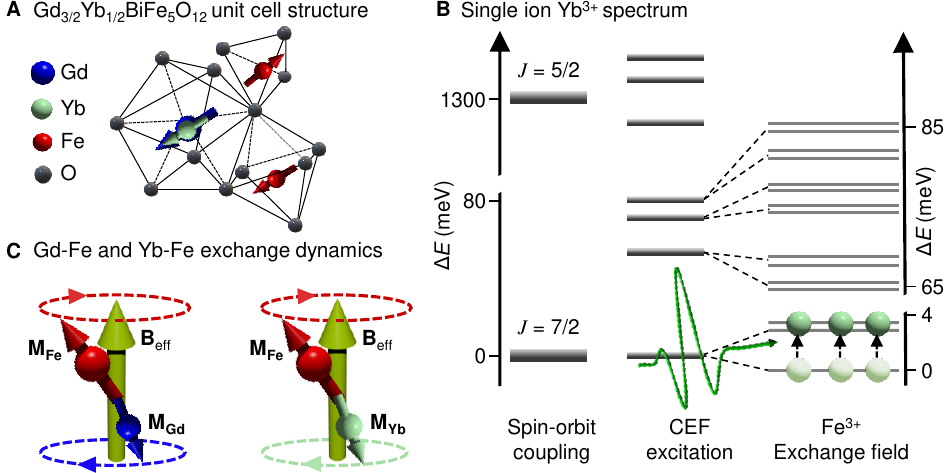}
 \caption{{\bf Crystal structure with sublattice exchange dynamics and Yb$^{3+}$ single ion excitation.} {\bf (A)} Unit cell structure of Gd$_{3/2}$Yb$_{1/2}$BiFe$_{5}$O$_{12}$, where the iron (Fe) ions occupy the tetrahedral and octahedral sites, and the rare-earth ions (Gd and Yb) are located at the dodecahedral site. The rare-earth magnetic moments couple to the net Fe magnetic moment via superexchange interactions mediated by O$^{2-}$ ions, thereby giving rise to the overall ferrimagnetic ordering. {\bf (B)} Electronic configuration of the 4$f^{13}$ electrons of the Yb$^{3+}$ ion, whose ground state is split due to strong spin-orbit coupling, crystal electric field, and Fe$^{3+}$-exchange field. The lowest-energy-level transitions are accessible within the measured THz spectral range. {\bf (C)} Schematic illustration of the precessional motion of Gd-Fe and Yb-Fe sublattice magnetization under different effective magnetic fields.}
\label{fig1}
\end{figure}

{\bf Experimental signatures of the interplay}\\
To investigate the synergy between the CEF transitions and sublattice magnetization dynamics, we performed temperature-dependent THz-TDS on Gd$_{3/2}$Yb$_{1/2}$BiFe$_{5}$O$_{12}$  across its magnetization compensation point, $T_{\rm MC}$. The single crystals are grown via liquid-phase epitaxy and are oriented along the (111) direction. As a room-temperature ferrimagnet, GdYb-BIG hosts distinct magnetic sublattices: iron ions occupy the tetrahedral and octahedral sites with antiparallel alignment, while RE ions (Gd and Yb) occupy the dodecahedral sites, see Figure~\ref{fig1}A. These RE ions couple to the net Fe moment through oxygen-mediated superexchange, establishing the ferrimagnetic ground state~\cite{Parchenko2013APL}. The system is characterized by a Curie temperature of 573 K and a compensation temperature of $T_{\rm MC} = 96$\,K~\cite{SatohNP2012}. Above $T_{\rm MC}$, the net magnetization is governed by the Fe sublattices; however, as the temperature drops, the RE contribution becomes dominant. Notably, the Yb sublattice moment emerges below 150 K and increases sharply at cryogenic temperatures, see Figure S1 of the Supporting Information. This provides us with a significant temperature window for studying the interplay between the CEF transitions and the collective magnetic order.

The Yb$^{3+}$ ion, further, exhibits a unique temperature dependence of its sublattice magnetization, characterized by the CEF excitations within the THz frequency range. Unlike the Gd$^{3+}$ (4$f^7$) ion, which possesses zero orbital angular momentum, the Yb$^{3+}$ ($4f^{13}$) ion features finite orbital ($L=3$) and spin ($S=1/2$) angular momenta. This configuration makes spin-orbit coupling the primary electronic interaction, splitting the $4f^{13}$ ground state into $J=7/2$ and $J=5/2$ multiplets. The surrounding crystalline electrostatic potential from ligand O$^{2-}$ ions further lifts the degeneracy of these multiplets into $J+1/2$ doublets. Additionally, a robust exchange field from the Fe sublattice acts on the Yb$^{3+}$ ions, attempting to align individual moments antiparallel to the net Fe magnetization~\cite{VivianePRB2022, SieversPRB1963}. This effective local field induces a final splitting of the doublets into low-energy states; notably, the lowest-energy spacing is resonant with the THz excitation (0.7\,THz) observed in our measurements (Figure~\ref{fig1}B). Consequently, the non-zero orbital angular momentum of Yb modifies the Yb-Fe anisotropy energy and its response to the Fe exchange field, leading to a distinct effective magnetic field that governs the coupled precessional dynamics of the sublattice magnetizations (Figure~\ref{fig1}C).

\begin{figure}[t!]
 \centering
 \includegraphics[width=0.85\textwidth]{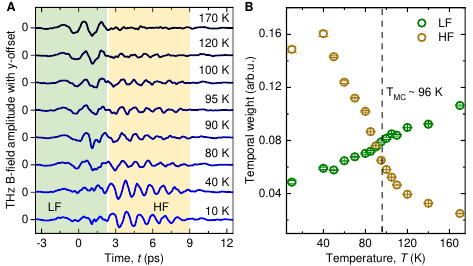}
 \caption{{\bf Temperature dependent time-transients and temporal weight analysis.} {\bf (A)} Temperature-dependent time transients of GdYb-BIG, recorded in the presence of an external in-plane magnetic field of 20\,mT. The time transients reveal low-frequency (LF) and high-frequency (HF) oscillatory components, denoted by the green- and yellow-shaded regions, respectively. As the LF component decreases, the HF component begins to develop below a certain temperature and strengthens thereafter. {\bf (B)} The temperature-dependent spectral weights associated with the LF and HF components cross near the magnetization compensation temperature (T$_{\rm MC}$), featuring a cross-talk characteristic to the coupled dynamics.}
\label{fig2}
\end{figure}

Figure~\ref{fig2}A presents the temperature-dependent THz time transients transmitted through a GdYb-BIG single crystal under a 20 mT in-plane external magnetic field (${\bf B}_{\rm ext}$). To isolate the field-induced magnetic response and suppress the significant background absorption inherent to the system in this frequency range, the transients recorded at ${\bf B}_{\rm ext} = 20$\,mT were subtracted from the zero-field measurements (see Supporting Information, Figure~S2). The differential signals reveal a pronounced temperature evolution characterized by the emergence of two distinct oscillatory components. In the high-temperature regime (e.g., 170 K), the response is dominated by a low-frequency (LF) mode, which gradually attenuates upon cooling. With temperatures approaching the magnetic compensation point ($T_{\rm MC}$), a high-frequency (HF) mode emerges and increases in amplitude as the temperature is further lowered. Strikingly, near $T_{\rm MC}$, the LF and HF components attain comparable spectral weights. This redistribution of dynamical weight is further quantified by the temporal weight analysis in Figure~\ref{fig2}B. The observed crossing of the LF and HF weights near $T_{\rm MC}$ provides strong evidence of coupled mode dynamics, signaling a fundamental reconfiguration of the system's magnetic excitation spectrum as it approaches the compensation threshold.

\begin{figure}[t!]
 \centering
 \includegraphics[width=0.85\textwidth]{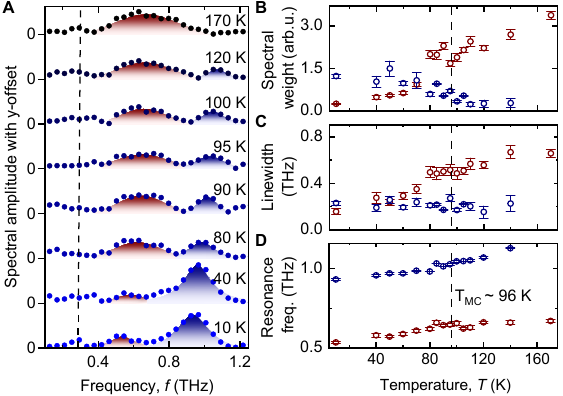}
 \caption{{\bf Temperature evolution of the sublattice exchange modes and the crystal electric field excitation.} {\bf (A)} Temperature-dependent spectra of the time transients in Figure~\ref{fig2}A. The spectra reveal the temperature evolution of the individual modes, namely, the Gd-Fe exchange dynamics (at 0.3\,THz), the Yb-ion CEF excitation (at 0.7\,THz), and the Yb-Fe exchange dynamics (at 1\,THz). The red- and blue-shaded regions represent the corresponding Lorentz fit of the CEF excitation and Yb-Fe exchange frequency components. {\bf (B)} The spectral weight of the CEF excitation and Yb-Fe exchange mode shows a redistribution near the magnetization compensation point. The temperature-dependent {\bf (C)} linewidth and {\bf (D)} the resonance frequency of these two modes, respectively. While the CEF linewidth narrows with decreasing temperature, the Yb-Fe exchange mode linewidth remains nearly unchanged. Moreover, both resonance modes exhibit a redshift in their frequencies as the temperature decreases.}
\label{fig3}
\end{figure}

To identify the frequency components of the LF and HF oscillations, we applied a Fast Fourier Transform (FFT) to the time-domain signals (Figure~\ref{fig2}A). The resulting temperature-dependent spectra in Figure~\ref{fig3}A reveal two prominent resonances centered at approximately 0.7\,THz and 1\,THz. The spectral amplitudes of these modes align well with those observed in the time domain. While the 0.7\,THz (LF) component is readily assigned to the CEF excitation of the Yb-single-ion, the emergence of the 1\,THz (HF) component was initially unexpected. Notably, the HF mode's intensity closely tracks the Yb sublattice magnetization; as temperature decreases, the increasing Yb magnetic moment suggests that this mode originates from Yb-Fe exchange dynamics, likely hybridized with the Yb-CEF excitation. Additionally, a weaker resonance appears near 0.3\,THz, which we tentatively attribute to the Gd-Fe exchange interactions.

While the 0.3\,THz mode remains nearly temperature-independent, the other two modes exhibit a pronounced temperature evolution. To analyze this behavior, the individual spectra were fitted using Lorentzian profiles (Figure~\ref{fig3}A). Notably, a significant redistribution of the spectral weight occurs between the CEF excitation and the Yb-Fe exchange mode near $T_{\rm MC}$, as shown in Figure~\ref{fig3}B. Furthermore, while the CEF excitation narrows upon cooling due to suppressed thermal fluctuations, the linewidth of the Yb-Fe exchange mode remains remarkably constant (Figure~\ref{fig3}C). This stability indicates a predominantly coherent precessional motion of the coupled sublattice magnetizations.

Notably, both the CEF and Yb-Fe exchange resonances exhibit a pronounced redshift upon cooling (Figure~\ref{fig3}D), directly contradicting the conventional blue-shift expected as exchange interactions strengthen at low temperatures. This systematic softening suggests a departure from standard exchange dynamics. We attribute this anomalous temperature dependence to a robust coupling between the Yb-ion CEF excitations and Yb-Fe exchange modes, mediated by strong spin-orbit interaction. Within this framework, the effective magnetic anisotropy of the Yb sublattice is renormalized by the competing influences of the internal Fe exchange field and the CEF manifold, ultimately driving the observed downward renormalization of the resonance frequencies.

\noindent{\bf Modeling the observed exchange renormalization}\\
To qualitatively elucidate the microscopic origin of the observed renormalization, we developed a theoretical framework predicated on the Landau-Lifshitz-Gilbert (LLG) equation~\cite{landau35, nowak2007handbook, Evans_2014, Skubic_2008}. The LLG equation governs the temporal evolution of the magnetization vector, providing a rigorous description of its precessional motion and the phenomenological damping. By employing this formalism, we can map the trajectories of magnetic moments and identify the underlying dissipation mechanisms driving the system's collective response. The dynamical evolution of the magnetization in our system is determined by the following LLG expression:
\begin{align}
\label{Eq1}
\frac{{\rm d}{\bf m}_{i}}{{\rm d}t} = -\frac{\gamma_i}{1+\alpha_i^2} \,{\bf m}_{i} \times \left[{\bf B}_{i}^{\rm eff} + \frac{{\alpha_i}}{\vert {\bf m}_{i}\vert}\left({\bf m}_i\times
   {\bf B}_{i}^{\rm eff}\right)\right],
\end{align}
where ${\bf m}_{i}$, $\gamma_i$, $\alpha_i$, and ${\bf B}_{i}^{\rm eff}$ denote the sublattice-dependent bulk magnetization, gyromagnetic ratio, Gilbert damping, and the effective magnetic field, respectively. The first term on the right side of the LLG equation signifies the precessional motion of the magnetization vector, whereas the second term explains the relaxation of that precessing magnetization towards its equilibrium direction. The effective field is calculated via the functional derivative $\mathbf{{B}}_{i}^{\rm eff} = -\delta \mathcal{F}/\delta \mathbf {m}_{i}$, where $\mathcal{F}$ is the free energy of the ferrimagnetic garnet system~\cite{Mondal2019PRB, Dutta2024, Mikuni_2025}. Because the system contains two distinct species of RE ions with unique evolution of magnetization with temperature (Figure~S1 of Supporting Information) and orbital angular momentum, a three-sublattice model is essential to capture the full magnetic behavior around the magnetization compensation temperature, $T_{\rm MC}$. Under this framework, the free energy is expressed as:
\begin{align}
\label{Eq2}
\mathcal{F}({\bf m}_{\rm Fe},{\bf m}_{\rm Gd},{\bf m}_{\rm Yb}) = \nonumber
    &- \lambda_{\rm Gd-Fe} \mathbf{m}_{\rm Gd} \cdot \mathbf{m}_{\rm Fe} - \lambda_{\rm Yb-Fe} \mathbf{m}_{\rm Yb} \cdot \mathbf{m}_{\rm Fe} - \lambda_{\rm Gd-Yb} \mathbf{m}_{\rm Gd} \cdot \mathbf{m}_{\rm Yb}  \nonumber \\
    &- \mathit K_{\rm Fe} \frac{\left(\mathbf{m}_{\rm Fe}\cdot \mathbf{n}\right)^{2}}{|\mathbf{m}_{\rm Fe}|^2} - \mathit K_{\rm Gd} \frac{\left(\mathbf{m}_{\rm Gd}\cdot \mathbf{n}\right)^{2}}{|\mathbf{m}_{\rm Gd}|^2} -\mathit K_{\rm Yb} \frac{\left(\mathbf{m}_{\rm Yb}\cdot \mathbf{n}\right)^{2}}{|\mathbf{m}_{\rm Yb}|^2} 
   \nonumber\\ &
    - [\mathbf{B}_{\rm THz} (t) + \mathbf{B}_{\rm ext}] \cdot(\mathbf{m}_{\rm Fe}+\mathbf{m}_{\rm Gd}+\mathbf{m}_{\rm Yb}) \nonumber\\ &
    + \frac{\mu_0}{2}(\mathbf{m}_{\rm Fe}\cdot \mathbf{n} + \mathbf{m}_{\rm Gd}\cdot \mathbf{n} + \mathbf{m}_{\rm Yb}\cdot \mathbf{n})^{2}. 
\end{align}
The first three terms on the right-hand side of Eq.~\eqref{Eq2} account for the exchange energy of the system, where $\lambda_{\rm Gd-Fe}$, $\lambda_{\rm Yb-Fe}$, and $\lambda_{\rm Gd-Yb}$ are the exchange constants mediating the inter-sublattice interactions among Fe, Gd, and Yb. The magneto-crystalline anisotropy energy is described by the terms involving $\mathit K_{\rm Fe}$, $\mathit K_{\rm Gd}$, and $\mathit K_{\rm Yb}$, which represent the anisotropy constants for the respective sublattices. The remaining terms in Eq.~\eqref{Eq2} correspond to the Zeeman and demagnetization energies. Here, $\mu_0$ denotes the permeability of free space, and $\mathbf{n}$ is the unit vector defining the easy axis of anisotropy, which is oriented along the $z$-direction. The system is driven by a time-dependent external THz magnetic pulse, $\mathbf {B}_{\rm THz}(t)$, expressed as: $\mathbf {B}_{\rm THz}(t) = \rm B_{0} \, \cos(2\pi f_{0}\Gamma [\mathrm {e}^{t/\Gamma}-1]) \; \mathrm {e}^{-t^{2}/\sigma^{2}}\; \hat{\mathbf{x}}$. This linearly polarized, chirped-pulse profile is chosen to closely match the experimental THz excitation conditions reported in Refs.~\cite{Dutta2025, Mukherjee2025PRM, Mukherjee2026PRB}. In this expression, $\rm B_{0}$ represents the pulse amplitude, $\mathrm{\sigma}$ is the temporal pulse width, $\rm {\Gamma}$ is the chirp parameter, and $\rm {f_0}$ denotes the frequency. The pulse is polarized along the $x$-direction, orthogonal to the equilibrium magnetization.

In GdYb-BIG, the high spin-orbit coupling, which results in higher Gilbert damping, necessitates the inclusion of an additional intrinsic spin-torque, the field-derivative torque (FDT), in the description of the system dynamics. This torque modifies the effective magnetic field of each sublattice $i$ by supplementing a time-derivative of the THz magnetic field as discussed in Ref~\cite{Dutta2025, Mukherjee2025PRM, Mukherjee2026PRB}. Note that the FDT was experimentally observed in the same ferrimagnetic system as we consider here. Such a time-derivative of the driving field is scaled by a pre-factor involving the unit cell volume per spin for each sublattice ($a^3_i$), gyromagnetic ratio ($\gamma_i$), Gilbert damping ($\alpha_{i}$), and the Bohr magneton ($\mu_{\rm B}$). For our numerical simulations, we utilized the following THz pulse parameters: $\rm B_0 = 20$ mT, $\rm \sigma = 1$ ps, $\rm f_0 = 0.6$ THz, and a chirp parameter $\rm \Gamma = 2.6$ ps that mimics the experimentally employed THz pulse. The Gilbert damping constants were set to $\alpha_{\rm Fe}= \alpha_{\rm Gd} = \alpha_{\rm Yb} =0.02$~\cite{Dutta2025}. While the gyromagnetic ratios for Fe and Gd were taken as $\gamma_{\rm Fe} = \gamma_{\rm Gd} = 1.76 \times 10^{11}$ s$^{-1}$T$^{-1}$, a reduced value of $\gamma_{\rm Yb} = 1.17 \times 10^{11}$ s$^{-1}$T$^{-1}$ was employed for the Yb sublattices to account for the significant orbital angular momentum contribution to its magnetization. The unit cell volumes per spin were defined as $a^3_{\rm Fe} = 1.2 \times 10^{-28}$ m$^3$ and $a^3_{\rm Gd} = a^3_{\rm Yb} = 8.5 \times 10^{-29}$ m$^3$, with the Bohr magneton $\mu_{\rm B}$ assuming standard values~\cite{Dutta2025}. Temperature-dependent sublattice magnetizations and anisotropy constants were incorporated based on the experimental observations from Refs.~\cite{Mikuni_2025, parchenko2016laser}. Using the above quantities, we solve three coupled LLG spin-dynamical equations following Eq.~\eqref{Eq1}, corresponding to the three magnetic sublattices, and compute the resulting magnetization properties.

\begin{figure}[t!]
 \centering
 \includegraphics[width=1.0\textwidth]{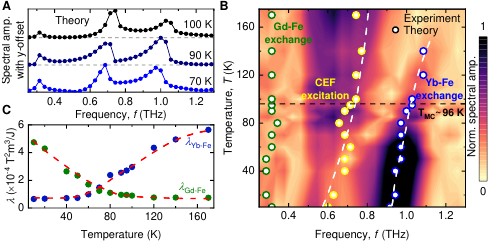}
 \vspace{-10pt}
 \caption{{\bf Hybridization between Yb-Fe exchange dynamics and Yb-ion crystal electric field excitation.} {\bf (A)} Theoretical resonance spectra at three representative temperatures, identifying the three primary modes: Gd-Fe exchange, Yb-ion crystal electric field (CEF) excitation, and Yb-Fe exchange. {\bf (B)} Contour plot of the temperature-dependent experimental spectra. The resonance frequencies derived from our theoretical model (open circles) exhibit excellent agreement with the experimental data, highlighting the hybridization between the Yb-ion CEF excitation and Yb-Fe exchange dynamics. White-dashed lines are provided as a guide to the eye. {\bf (C)} Temperature dependence of the Yb-Fe and Gd-Fe exchange energy. The reduction in Yb-Fe exchange energy at lower temperatures drives a pronounced red-shift in both the Yb-ion CEF and Yb-Fe exchange resonance modes. The red-dashed line is a guide to the eye.}
\label{fig4}
\end{figure}

In this three-sublattice model, the magnetization dynamics are governed by competing inter-sublattice exchange interactions: ferromagnetic Gd-Yb coupling and antiferromagnetic Gd-Fe and Yb-Fe couplings~\cite{Gorbatov2021, Xie2017, Wang2020}. Although the Gd-Yb exchange magnitude is significantly smaller than the others, it exerts a discernible influence on the global frequency response~\cite{Jensen1991}. The observed spectra (Figure~\ref{fig4}A) represent a cumulative response of all three interactions, though their spectral weights vary across the frequency domain. Specifically, the 0.3\,THz resonance is dominated by Gd-Fe exchange, while the 0.7\,THz and 1\,THz peaks are primarily driven by Yb-Fe interactions. The theoretical resonance frequencies associated with Gd-Fe exchange mode, Yb single-ion CEF excitations, and Yb-Fe exchange mode show good agreement with the experimentally observed temperature-dependent spectra (Figure~\ref{fig4}B). Physically, the Gd-Fe exchange energy increases upon cooling, consistent with the strengthening of magnetic order at lower temperatures. Conversely, the Yb-Fe exchange energy decreases with decreasing temperature, inducing a characteristic redshift in the 0.7\,THz and 1\,THz peaks. This reduction in Yb-Fe coupling also moderates the 0.3\,THz peak. Although this low-frequency mode is primarily governed by the increasing Gd–Fe exchange interaction, its frequency remains relatively stable due to the countervailing influence of the Yb-Fe sublattice. This cross-sublattice coupling further introduces a redshift of the high-frequency peaks. 

The spin-orbit coupling and non-zero orbital angular momentum of the Yb$^{3+}$ ion facilitate the lifting of ground-state degeneracy via the CEF. Near room temperature, thermal agitation ensures a nearly uniform population across these CEF-split levels, rendering the bulk magnetization of the Yb sublattice negligible. However, as the temperature decreases, the preferential occupation of lower-energy CEF states significantly enhances the net Yb sublattice magnetization. This temperature-dependent redistribution of states renders the Yb$^{3+}$ ions energetically non-equivalent to neighboring Fe$^{3+}$ ions, inducing a highly anisotropic exchange interaction ($\lambda_{\rm Yb-Fe}$) defined as~\cite{VivianePRB2022, WickersheimPR1961, WickersheimPRL1960,Harris1963}
\begin{equation}
\lambda_{\rm Yb-Fe} =
\begin{pmatrix}
2.26\lambda & 3.30\lambda & -0.47\lambda \\
-1.91\lambda & -2.87\lambda & 3.22\lambda \\
4.04\lambda & 4.68\lambda & -2.24\lambda
\end{pmatrix},
\label{Eq3}
\end{equation}
where $\lambda$ is a temperature-dependent variable whose values change in accordance with Figure~\ref{fig4}C (see S4 of Supporting Information for detailed simulations implementation). The emergence of distinct dynamical signatures tracks this quantum-mechanical evolution: a 0.7\,THz resonance peak becomes discernible at 170\,K, driven by the onset of anisotropic exchange, followed by a 1\,THz peak at 140\,K as the Yb ground-state magnetization becomes appreciable. Below these thresholds, the suppression of thermal fluctuations further consolidates the population within the lowest CEF doublets~\cite{Iida1967magnetostriction}. This results in a reduction of anisotropy in the exchange interaction between the Yb and Fe sublattices, accompanied by an increase in the bulk ground-state magnetization of the Yb sublattice, which is manifested as a shift of spectral weight between the 0.7\,THz and 1\,THz exchange modes. As the anisotropy in the Yb-Fe exchange interaction gets lower, the collective value of Yb-Fe exchange energy decreases as the temperature decreases. This decrease in the Yb-Fe exchange, in turn, results in a redshift of its dominant frequency modes. In contrast, the Yb-Fe exchange interaction exerts only an indirect influence on the 0.3\,THz frequency peak and is therefore unable to induce a red shift in this mode. However, it suppresses the increase in this peak, even though the Gd-Fe exchange interaction strengthens at lower temperatures.

\section*{CONCLUSION}
In summary, our temperature-dependent THz-TDS study reveals a robust hybridization between localized Yb-ion CEF excitations and collective Yb–Fe exchange-driven spin dynamics in the ferrimagnetic garnet Gd$_{3/2}$Yb$_{1/2}$BiFe$_{5}$O$_{12}$. A marked redistribution of temporal and spectral weight near the compensation point signals the emergence of strongly coupled low-energy dynamics. Crucially, the Yb-Fe exchange mode undergoes an anomalous redshift upon cooling, diverging from the conventional hardening typically driven by enhanced anisotropic exchange interactions at low temperatures. We attribute this unconventional softening to a renormalization of the exchange energy, arising from the competing interplay between the iron exchange field and Yb-ion CEF excitations -- a process mediated by spin-orbit coupling. Beyond these findings, we propose a generalized material design principle in which the THz spin dynamics of rare-earth garnets can be engineered by selecting and tuning the rare-earth sublattices. Rare-earth ions with non-zero orbital angular momentum (Yb$^{3+}$, Ho$^{3+}$, Er$^{3+}$, and Tm$^{3+}$) can introduce CEF-mediated hybridization that fundamentally alters the exchange mode spectrum, which is inaccessible for the other S-state rare-earth (Y$^{3+}$, Gd$^{3+}$) garnets. The strength of hybridization, the mode frequencies, and the temperature at which spectral weight transfer occurs are all tunable via the rare-earth elements and their concentrations. This opens a pathway to tailor exchange-controlled ultrafast dynamics in rare-earth iron garnets for future spintronic and magnonic architectures.

\section*{METHODS}

{\bf Sample characteristics:} The single crystals of Gd$_{3/2}$Yb$_{1/2}$BiFe$_{5}$O$_{12}$ are grown by the liquid phase epitaxy method and oriented along the [111] direction. The thickness of the crystals used in our experiment is 380\,$\mu$m. The crystal exhibits a Curie temperature of 573\,K and magnetization compensation temperature of 96\,K. Doping the garnet with rare-earth ions promotes spin precession at THz frequencies near room temperature.

\noindent{\bf THz time-domain spectroscopy:} A Ti:Sapphire laser operating at a wavelength of 800\,nm, pulse duration 100\,fs, repetition rate 1\,kHz, and pulse energy of 3.5\,mJ/pulse, is used to generate single-cycle THz pulses via optical rectification in a 0.5\,mm thick (110)-oriented ZnTe crystal. Approximately 90\% of the fundamental beam is used for the THz generation, while the remaining 10\% is used as the gating beam for the free-space electro-optic sampling. The generated THz pulses are guided onto the sample using a set of off-axis parabolic mirrors. Subsequently, the transmitted THz light and the gating beam are collinearly focused onto a 1.4\,mm thick (110)-oriented ZnTe crystal, which is used for electro-optic detection. The THz light-induced birefringence in the ZnTe crystal results in an ellipticity of the gating beam, which is measured using a quarter-wave plate, a Wollaston prism, and a balanced photo-diode. During the measurements, an external in-plane magnetic field of ${\rm \bf B}_{\rm ext} = 20$\,mT is applied perpendicular to the THz magnetic field (${\rm \bf B}_{\rm THz}$). All temperature-dependent measurements are carried out in an inert nitrogen atmosphere to prevent water absorption by the THz light.

\clearpage

\begin{addendum}
\item[Correspondence:] All correspondence should be directed to Ritwik Mondal (email: ritwik@iitism.ac.in) and Shovon Pal (shovon.pal@niser.ac.in).

\item[Data and code availability:] Data and the codes supporting this study are available from the corresponding authors upon reasonable request.

\item[Author Contributions] All authors contributed to the discussion and interpretation of the experiment and to the completion of the manuscript. A.D. and P.M. contributed equally to this work. A.D. performed the experiments. A.D. and P.M. performed the data analysis. P.M. and R.M. developed the theoretical model. A.D. and S.P. conceived the project while S.P. supervised the experiments. A.D., P.M., R.M., and S.P. drafted the manuscript.

\item[Acknowledgments] A.D. and S.P. acknowledge the support from DAE through the projects RIN4001 and RNI4011. S.P. additionally acknowledges the start-up support from DAE through NISER. S.P. also acknowledges the support of the International Collaborative Research Program of the Institute for Chemical Research, Kyoto University (grant no.~2026-6). P.M. and  R.M. acknowledge funding support from the SERB-SRG via Project No. SRG/2023/000612 and the faculty research scheme at IIT (ISM) Dhanbad, India, under Project No. FRS(196)/2023-2024/PHYSICS. The authors acknowledge Takuya Satoh and Hideki Hirori for the fruitful discussions.

\item[Competing Interests:] The authors declare that they have no competing financial interests.
\end{addendum}

\end{document}


\maketitle
\begin{affiliations}
\item School of Physical Sciences, National Institute of Science Education and Research, Jatni, 752 050 Odisha, India
\item Homi Bhaba National Institute, Training School Complex, Anushakti Nagar, 400 094 Mumbai, India
\item Department of Physics, Indian Institute of Technology (ISM) Dhanbad, 826 004 Dhanbad, India
\end{affiliations}

\begin{abstract}
This supplementary information contains further details on temperature-dependent sublattice magnetization, THz time transients in presence and absence of the external magnetic field, the temperature-dependent experimental and theoretical spectra and simulation for anisotropic exchange interaction. The contents are sectionized as:
\begin{description}
    \item S1. Temperature-dependent sublattice magnetization
    \item S2. Time transients in presence and absence of external magnetic field
    \item S3. Temperature dependent experimental and theoretical spectra
    \item S4. Simulation details for anisotropic exchange interaction
\end{description}

\end{abstract}

\maketitle

\clearpage

\section{Temperature-dependent sublattice magnetization}
\begin{figure}[h!]
 \centering
 \includegraphics[width=0.5\textwidth]{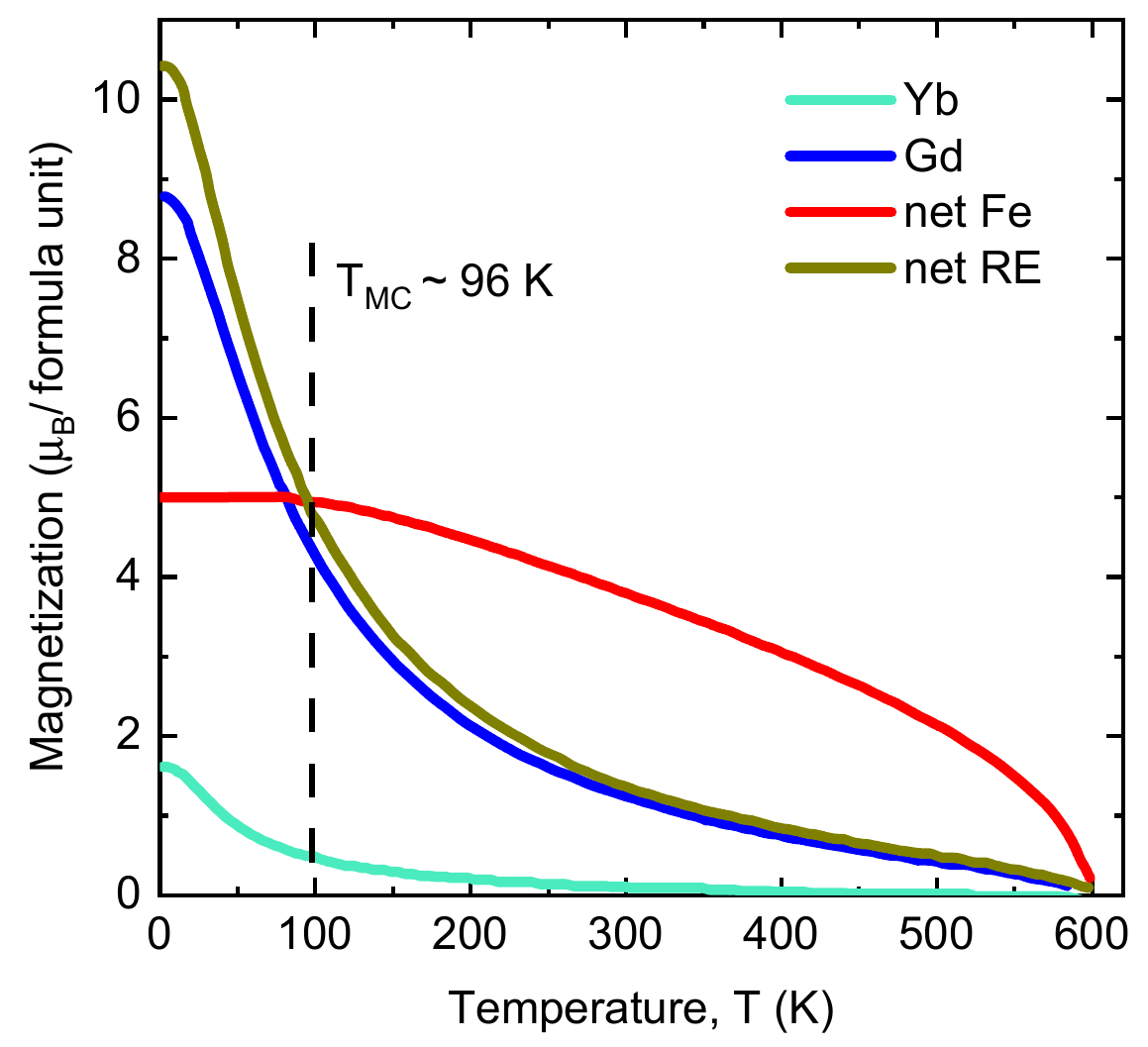}
 \caption{Temperature-dependent sublattice magnetization of Gd$_{3/2}$Yb$_{1/2}$BiFe$_{5}$O$_{12}$ (adapted from Figure~3.15 of Ref.~\cite{parchenko2016laser}). At magnetization compensation temperature ($T_{\rm MC}$), the net rare-earth (RE) moment cancels the net Fe moment.}
 \label{figS1}
\end{figure}

Figure~\ref{figS1} presents the temperature-dependent sublattice magnetizations of Gd$_{3/2}$Yb$_{1/2}$BiFe$_{5}$O$_{12}$, calculated by molecular field theory. At room temperature, the net rare-earth (RE) magnetic moment is predominantly governed by Gd ions, while the Yb magnetic moment gradually develops and becomes significant below approximately 150\,K. Its contribution further increases below the compensation temperature ($T_{\rm MC}$), where the net RE and Fe sublattice moments become equal in magnitude and opposite in direction. Notably, the temperature evolution of the Yb magnetic moment closely follows the behavior of the Yb-Fe exchange interaction mode near 1\,THz, highlighting a strong correlation between the Yb sublattice magnetization and low-energy exchange dynamics. 

\section{Time transients in presence and absence of external magnetic field}

\begin{figure}[t!]
 \centering
 \includegraphics[width=1.0\textwidth]{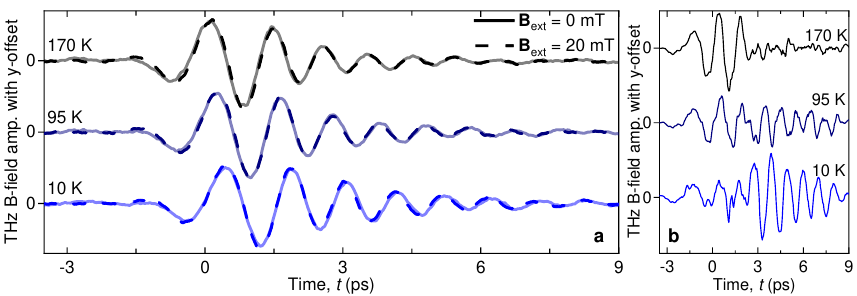}
 \caption{{\bf(a)} Temperature-dependent time transients of Gd$_{3/2}$Yb$_{1/2}$BiFe$_{5}$O$_{12}$, recorded in the absence (dashed lines) and presence (solid lines) of an external in-plane magnetic field of 20\,mT. The oscillation in the time transients at ${\bf B}_{\rm ext}$ = 20\,mT is attributed to the inherent material absorption within the measured THz frequency range. {\bf(b)} The subtracted signal as a function of temperature, featuring a low-frequency and high-frequency oscillation which shows pronounced and distinct temperature evolution.}
\label{figS2}
\end{figure}

Figure~\ref{figS2}a presents the temperature-dependent THz time transients of Gd$_{3/2}$Yb$_{1/2}$BiFe$_{5}$O$_{12}$ measured in the presence and absence of an external in-plane magnetic field of 20\,mT (${\bf B}_{\rm ext}$). As evident from Figure~\ref{figS2}a, the transients, recorded without ${\bf B}_{\rm ext}$, exhibit prominent oscillatory features, showing a slight temporal shift relative to the time transients, recoded in the presence of ${\bf B}_{\rm ext}$ = 20\,mT. We attribute these background oscillations to the intrinsic absorption of the material within the measured THz frequency range~\cite{Dutta2025}. To isolate the magnetic field induced response and to suppress the response from the background absorption, the zero-field transients are subtracted from the corresponding field measurements at each temperature. The resulting differential time-domain signals are shown in Figure S2b, where distinct low-frequency (LF) and high-frequency (HF) oscillations emerge and evolve systematically with temperature. 

\section{Temperature-dependent experimental and theoretical spectra}

\begin{figure}[h!]
 \centering
 \includegraphics[width=0.85\textwidth]{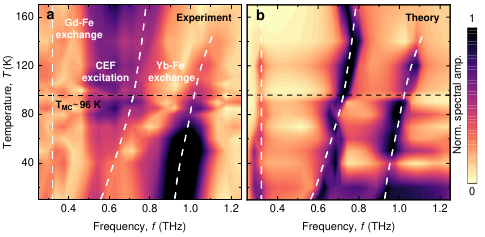}
 \caption{{\bf(a)}, {\bf(b)} Temperature-dependent 2D contour plot of the experimental and theoretically obtained spectra of GdYb-BIG, respectively. The theoretical spectra, obtained from the three-sublattice model, corroborate the experimental results and reveal a strong hybridization between the localized Yb-ion crystal-electric-field (CEF) excitation and the Yb-Fe exchange interaction. The white dashed lines are the guide-to-the-eye, highlighting the temperature evolution of the coupled resonances.}
\label{figS3}
\end{figure}

Figure~\ref{figS3} a,b presents the temperature-dependent experimental and theoretically obtained spectra corresponding to the Gd-Fe exchange dynamics and the coupled Yb-ion crystal electric field (CEF) and Yb-Fe exchange excitations. Within the framework of the three-sublattice model (as discussed in detail in the main article), the calculated spectra reproduce the experimental observations remarkably well, featuring the coupled low-energy spin dynamics across the investigated temperature range. In particular, the simulations clearly reveal a strong hybridization between the localized Yb-ion crystal electric field (CEF) excitation and the collective Yb-Fe exchange mode, which is manifested as a redshift in the resonance frequency upon cooling, in excellent agreement with the experimental results. 

\section{Simulation details for anisotropic exchange interaction}
The hybridization of the frequency modes near the magnetic compensation temperature is governed by the interplay among the three inter-sublattice exchange interactions. Of these, the Gd-Yb and Gd-Fe exchange interactions are predominantly isotropic in nature, whereas the Yb-Fe exchange interaction exhibits significant anisotropy. In general, the exchange interaction between two magnetic moments can be expressed in matrix form as
\[
\lambda =
\begin{pmatrix}
\lambda_{11} & \lambda_{12} & \lambda_{13} \\
\lambda_{21} & \lambda_{22} & \lambda_{23} \\
\lambda_{31} & \lambda_{32} & \lambda_{33}
\end{pmatrix}
\]

For isotropic exchange interaction, the off-diagonal terms are zero, and the diagonal elements are the same $\lambda_{11} = \lambda_{22} = \lambda_{33}$. In contrast, for anisotropic exchange interaction, all the matrix elements of $\lambda$ have non-zero values, as shown in Eq.(3) in the main text. The inclusion of isotropic exchange interactions in the effective magnetic field is relatively straightforward. However, the contribution arising from anisotropic exchange interactions requires a more careful treatment. In this section of the Supporting Information, we explicitly derive the component-wise form of the effective magnetic field,  $\mathbf{B}_i^{\rm eff}$, for the Yb and Fe sublattices in the presence of anisotropic Yb-Fe exchange coupling. Effective field components for the Fe sublattice,
\begin{align}
\mathbf{B}_{x,\mathrm{Fe}}^{\rm eff} &= 
\lambda_{\mathrm{Gd-Fe}}\, m_{x,\mathrm{Gd}}
+ \lambda_{11}\, m_{x,\mathrm{Yb}}
+ \lambda_{12}\, m_{y,\mathrm{Yb}}
+ \lambda_{13}\, m_{z,\mathrm{Yb}}
+ \mathrm{B_{\rm x, ext}} \nonumber \\  &
+ \mathbf{B_{\rm THz}}(t)
- \left(\frac{\alpha_{\rm Fe} a^{3}_{\mathrm{Fe}}}{\gamma_{\rm Fe} \mu_B \mu_0}\right)
\frac{{\rm d}\mathbf{B_{\rm THz}}}{{\rm d}t},
\\[8pt]
\mathbf{B}_{y,\mathrm{Fe}}^{\rm eff} &= 
\lambda_{\mathrm{Gd-Fe}}\, m_{y,\mathrm{Gd}}
+ \lambda_{22}\, m_{y,\mathrm{Yb}}
+ \lambda_{21}\, m_{x,\mathrm{Yb}}
+ \lambda_{23}\, m_{z,\mathrm{Yb}}
+ \mathrm{B_{\rm y, ext}},
\\[8pt]
\mathbf{B}_{z,\mathrm{Fe}}^{\rm eff} &= 
\lambda_{\mathrm{Gd-Fe}}\, m_{z,\mathrm{Gd}}
+ \lambda_{33}\, m_{z,\mathrm{Yb}}
+ \lambda_{31}\, m_{x,\mathrm{Yb}}
+ \lambda_{32}\, m_{y,\mathrm{Yb}}
+ \mathrm{B_{\rm z, ext}} \nonumber \\  &
+ \frac{2K_{\mathrm{Fe}}\, m_{z,\mathrm{Fe}}}
       {m_{x,\mathrm{Fe}}^2 + m_{y,\mathrm{Fe}}^2 + m_{z,\mathrm{Fe}}^2}
- \mu_0
\left(
m_{z,\mathrm{Gd}} + m_{z,\mathrm{Fe}} + m_{z,\mathrm{Yb}}
\right).
\end{align}
Effective field components for the Yb sublattice,
\begin{align}
\mathbf{B}_{x,\mathrm{Yb}}^{\rm eff} &= 
\lambda_{\mathrm{Gd-Yb}}\, m_{x,\mathrm{Gd}}
+ \lambda_{11}\, m_{x,\mathrm{Fe}}
+ \lambda_{21}\, m_{y,\mathrm{Fe}}
+ \lambda_{31}\, m_{z,\mathrm{Fe}}
+\mathrm{B_{\rm x, ext}} \nonumber \\ &
+ \mathbf{B_{\rm THz}}(t)
- \left(\frac{\alpha_{\rm Yb} a^{3}_{,\mathrm{Yb}}}{\gamma_{\rm Yb} \mu_B \mu_0}\right)
\frac{{\rm d}\mathbf{B_{\rm THz}}}{{\rm d}t},
\\[8pt]
\mathbf{B}_{y,\mathrm{Yb}}^{\rm eff} &= 
\lambda_{\mathrm{Gd-Yb}}\, m_{y,\mathrm{Gd}}
+ \lambda_{22}\, m_{y,\mathrm{Fe}}
+ \lambda_{12}\, m_{x,\mathrm{Fe}}
+ \lambda_{32}\, m_{z,\mathrm{Fe}}
+ \mathrm{B_{\rm y, ext}},
\\[8pt]
\mathbf{B}_{z,\mathrm{Yb}}^{\rm eff} &= 
\lambda_{\mathrm{Gd-Yb}}\, m_{z,\mathrm{Gd}}
+ \lambda_{33}\, m_{z,\mathrm{Fe}}
+ \lambda_{23}\, m_{y,\mathrm{Fe}}
+ \lambda_{13}\, m_{x,\mathrm{Fe}}
+ \mathrm{B_{\rm z, ext}} \nonumber \\ &
+ \frac{2K_{\mathrm{Yb}}\, m_{z,\mathrm{Yb}}}
       {m_{x,\mathrm{Yb}}^2 + m_{y,\mathrm{Yb}}^2 + m_{z,\mathrm{Yb}}^2}
- \mu_0
\left(
m_{z,\mathrm{Gd}} + m_{z,\mathrm{Fe}} + m_{z,\mathrm{Yb}}
\right).
\end{align}
Here, the matrix components of $\lambda$ should be understood as the components of the anisotropic Yb-Fe exchange interaction. We incorporate this effective field into the three-sublattice LLG spin-dynamical equations and solve them numerically to determine the resonance frequencies. The resulting theoretical simulations show excellent agreement with the experimentally observed hybridization of the frequency modes. 

When the diagonal elements of the Yb-Fe exchange matrix are set to zero while retaining finite off-diagonal elements, the higher-frequency modes exhibit a pronounced redshift, whereas the 0.3,THz mode undergoes only a weak redshift. This behavior originates from the reduction of the effective Yb-Fe exchange energy, while the exchange anisotropy remains preserved through the finite off-diagonal components of the exchange matrix.

In contrast, when the off-diagonal elements are set to zero while retaining only the diagonal components, the spectrum loses its characteristic structure and fragments into several smaller peaks. This behavior signifies the suppression of the CEF-induced anisotropy in the Yb-Fe exchange interaction, highlighting the crucial role of the off-diagonal exchange components in sustaining the anisotropic coupling and the collective hybridized nature of the resonance modes.